\documentclass[twocolumn, superscriptaddress, preprintnumbers, amsmath, amssymb, aps]{revtex4}

\usepackage{graphicx}
\usepackage[caption=false]{subfig}
\usepackage{amsthm}
\usepackage{amsmath}
\usepackage{tensor}
\usepackage{color}
\usepackage{bm}
\usepackage[all]{xy}
\usepackage{tikz}
\usepackage{dsfont}
\usepackage{times,txfonts}
\usepackage{algorithm}
\usepackage{algpseudocode}
\usetikzlibrary{positioning}
\usepackage{braket}
\usepackage{enumitem}
\usepackage{booktabs}
\usepackage{makecell}
\usepackage{multirow}
\usepackage{diagbox}
\usepackage{graphicx}
\renewcommand{\arraystretch}{2}
\theoremstyle{definition}

\usepackage{xcolor}
\newcommand{\rev}[1]{#1}

\begin{document}

\title{Locally Acting Grover Mixers for Constraint-Preserving QAOA}

\author{Minjin Choi}
% \email{mathcmj89@gmail.com}
\affiliation{
Center for Quantum Information R\&D, Korea Institute of Science and Technology Information~(KISTI), Daejeon 34141, Republic of Korea
}

\author{Dongkeun Lee}
\affiliation{
Center for Quantum Information R\&D, Korea Institute of Science and Technology Information~(KISTI), Daejeon 34141, Republic of Korea
}
\affiliation{
Division of Quantum Information, KISTI School, Korea University of Science and Technology, Daejeon 34141, Republic of Korea
}

\author{Junghee Ryu}
\email{junghee@kisti.re.kr}
\affiliation{
Center for Quantum Information R\&D, Korea Institute of Science and Technology Information~(KISTI), Daejeon 34141, Republic of Korea
}
\affiliation{
Division of Quantum Information, KISTI School, Korea University of Science and Technology, Daejeon 34141, Republic of Korea
}

\date{\today}

%%%%%%%%%%%%%%%%%%%%%%%%%%%%%%%%%%%%%%%%%%%%%%%%%%%%%%%%%%%%%%%%%%%%
%%%
%%% Abstract
%%%
%%%%%%%%%%%%%%%%%%%%%%%%%%%%%%%%%%%%%%%%%%%%%%%%%%%%%%%%%%%%%%%%%%%%
\begin{abstract}
The Grover mixer quantum alternating operator ansatz~(GM-QAOA) employs the Grover mixer to confine the quantum evolution to the feasible subspace defined by the problem.
Its mixing unitary, however, requires a global multi-controlled phase-shift gate acting on all qubits, resulting in substantial circuit overhead on near-term quantum devices.
In this work, we propose locally acting Grover mixers \rev{with independent variational parameters,}
tailored to initial states that admit a product structure over disjoint qubit subsystems, which may be obtained by encoding only a subset of problem constraints into the initial state preparation.
The proposed method preserves the search space defined by the initial state while \rev{reducing the mixing-circuit gate count and depth relative to the global Grover mixer,} as the global multi-controlled phase-shift gate is replaced with local operations on disjoint subsystems.
Numerical simulations on the exact-cover problem and the traveling salesman problem~(TSP) demonstrate that the proposed method achieves \rev{solution probabilities comparable to those} of the original GM-QAOA, while using shallower circuits with fewer gates.
\rev{
Comparisons with shared-angle local Grover and partitioned XY-ring mixers characterize the trade-offs between solution quality, circuit complexity, and optimization overhead.
}
We further compare two constraint encoding strategies for the TSP, encoding only a subset of constraints versus all constraints into the initial state preparation, and show that the former combined with the proposed mixer \rev{can yield} markedly more compact circuits at the point where comparable solution quality is achieved.
\end{abstract}

\maketitle

%%%%%%%%%%%%%%%%%%%%%%%%%%%%%%%%%%%%%%%%%%%%%%%%%%%%%%%%%%%%%%%%%%%%
%%%
%%% Introduction
%%%
%%%%%%%%%%%%%%%%%%%%%%%%%%%%%%%%%%%%%%%%%%%%%%%%%%%%%%%%%%%%%%%%%%%%
\section{Introduction}
\label{sec:introduction}

Quantum computing has the potential to address combinatorial optimization problems more efficiently than classical methods, and algorithms such as Grover's search~\cite{Grover1996} and Grover adaptive search~\cite{Baritompa2005, Gilliam2021} offer quantum speedups for optimization tasks.
However, their realization requires fault-tolerant quantum computing, which remains beyond the reach of current quantum hardware.
Instead, the currently available devices are noisy intermediate-scale quantum devices~\cite{Preskill2018}, characterized by significant noise and limited circuit depth.
This has motivated the development of variational quantum algorithms designed to work within these restrictions~\cite{Cerezo2021}.
Among such near-term approaches, the quantum approximate optimization algorithm~(standard QAOA), introduced by Farhi \textit{et al.}~\cite{Farhi2014}, has attracted considerable attention.
It constructs a parameterized quantum circuit through alternating phase-separation and mixing unitaries, with variational parameters optimized classically.
This algorithm was later generalized to the quantum alternating operator ansatz~(QAOA)~\cite{Hadfield2019}, which allows these unitaries to be flexibly constructed to respect the structure of the optimization problem.

A central challenge in applying QAOA to constrained optimization problems is that even if constraints are encoded in the initial state, the standard mixer does not preserve the feasible subspace and can drive the quantum state outside it.
To address this, various constraint-preserving mixers have been developed, including the XY-mixer~\cite{Wang2020} and more general feasibility-preserving constructions~\rev{\cite{Fuchs2022, Fuchs2024}}.
Among these approaches, Grover mixer QAOA~(GM-QAOA), proposed by B{\"a}rtschi and Eidenbenz~\cite{Bartschi2020}, provides a particularly appealing solution.
In GM-QAOA, the mixing operator is a parameterized unitary whose generator is the projector onto the initial state, thereby automatically preserving the feasible subspace throughout the optimization.

GM-QAOA has been demonstrated on a range of combinatorial optimization problems, including the traveling salesman problem~(TSP), the $k$-vertex cover problem, and discrete portfolio rebalancing~\cite{Bartschi2020}. 
It has since been applied to problems such as the capacitated vehicle routing problem~\cite{Xie2024}, satisfiability problems~\cite{Zhang2025, Seo2025}, and solution counting~\cite{Drapeau2025}.
\rev{Grover mixers have also been investigated in different encoding schemes for the weighted MAX $k$-CUT problem, including constructions that apply Grover mixing locally to individual registers~\cite{Fuchs2025}.}
Extensions of the framework have been explored, including the Grover mixer threshold QAOA~\cite{Golden2021}, which replaces the phase-separation operator with a threshold function and establishes a closer connection to Grover's search algorithm, as well as hybrid approaches incorporating Grover mixers alongside other mixing strategies~\cite{Kim2023}.
On the theoretical side, the analytical properties of GM-QAOA and its variants have been studied~\cite{Bridi2024, Xie2025}, and provable avoidance of barren plateaus has been demonstrated~\cite{Tsvelikhovskiy2025}.
Efficient preparation of the required initial states has also been investigated~\cite{Bartschi2019, Bartschi2022}.
Collectively, these results highlight GM-QAOA as a well-studied and promising framework for constrained quantum optimization.

\rev{In the original GM-QAOA construction, however,} the global multi-controlled phase-shift gate that constitutes the mixing unitary is costly to implement in practice.
Since near-term quantum devices natively support only one- and two-qubit operations, such a gate must be decomposed into elementary gates~\cite{Barenco1995, Nielsen2010}.
Although efficient decomposition methods have been actively studied~\cite{Claudon2024, Zindorf2025}, implementing an $n$-qubit multi-controlled gate still requires a \rev{substantial number of elementary gates that increases with $n$,} posing serious challenges for near-term quantum devices.
This motivates the question of whether the structure of the feasible subspace can be exploited to design a more circuit-efficient mixer.
\rev{One approach is to exploit a product structure in the initial state and apply Grover mixing locally to individual subsystems, as considered in Ref.~\cite{Fuchs2025}.}

In this work, we propose locally acting Grover mixers \rev{with independent variational parameters for each subsystem,} tailored to initial states that admit a product structure over disjoint qubit subsystems.
\rev{This differs from Ref.~\cite{Fuchs2025}, where a single mixing angle is shared across subsystems within each layer.}
By confining each mixer to act independently within its respective subsystem, the proposed method substantially reduces circuit complexity while preserving the search space.
We demonstrate this approach on \rev{exact-cover and TSP instances requiring up to 10 qubits}, showing that the proposed mixer achieves \rev{solution probabilities comparable to those} of the original GM-QAOA with significantly reduced gate count and circuit depth.
\rev{Additional linear-ramp simulations~\cite{Montanez2025} extend the TSP evaluation to instances requiring up to 25 qubits.
Noise simulations of the full circuit with gate-level depolarizing noise also support improved robustness to gate noise compared with GM-QAOA.}
We also examine two constraint encoding strategies for the TSP, namely encoding all constraints into the initial state preparation versus encoding only a subset, where the latter yields a product-structured initial state amenable to the proposed mixer.
We show that partial encoding combined with the proposed mixer \rev{can lead} to considerably more compact circuits at the point where comparable solution quality is achieved.
\rev{Comparisons with shared-angle local Grover and partitioned XY-ring mixers clarify the role of independent mixing parameters and the trade-offs between solution quality, circuit complexity, and optimization overhead.}

The remainder of the paper is organized as follows.
Section~\ref{sec:methods} reviews the GM-QAOA framework and introduces the proposed locally acting Grover mixers.
Section~\ref{sec:results} presents numerical results and resource comparisons.
Section~\ref{sec:conclusion} concludes with a summary and directions for future work.

%%%%%%%%%%%%%%%%%%%%%%%%%%%%%%%%%%%%%%%%%%%%%%%%%%%%%%%%%%%%%%%%%%%%
%%%
%%% Methods
%%%
%%%%%%%%%%%%%%%%%%%%%%%%%%%%%%%%%%%%%%%%%%%%%%%%%%%%%%%%%%%%%%%%%%%%
\section{Methods}
\label{sec:methods}

\subsection{QAOA with the Grover Mixer}
\label{subsec.II.1}
We begin with a brief review of GM-QAOA~\cite{Bartschi2020}, which builds on the QAOA framework~\cite{Hadfield2019}.
The QAOA framework extends the standard QAOA~\cite{Farhi2014} by allowing more flexible constructions of phase-separation and mixing operators through parameterized families of unitaries, rather than restricting them to time evolutions under fixed Hamiltonians.
This flexibility enables the design of mixers that preserve a prescribed search space associated with the problem constraints.
Among such constructions, GM-QAOA employs the Grover mixer, whose generator is the projector onto the initial state in analogy with the diffusion operator in Grover's search, to preserve the associated search space.

\begin{figure}[t]
\centering
\includegraphics[width=0.47\textwidth]{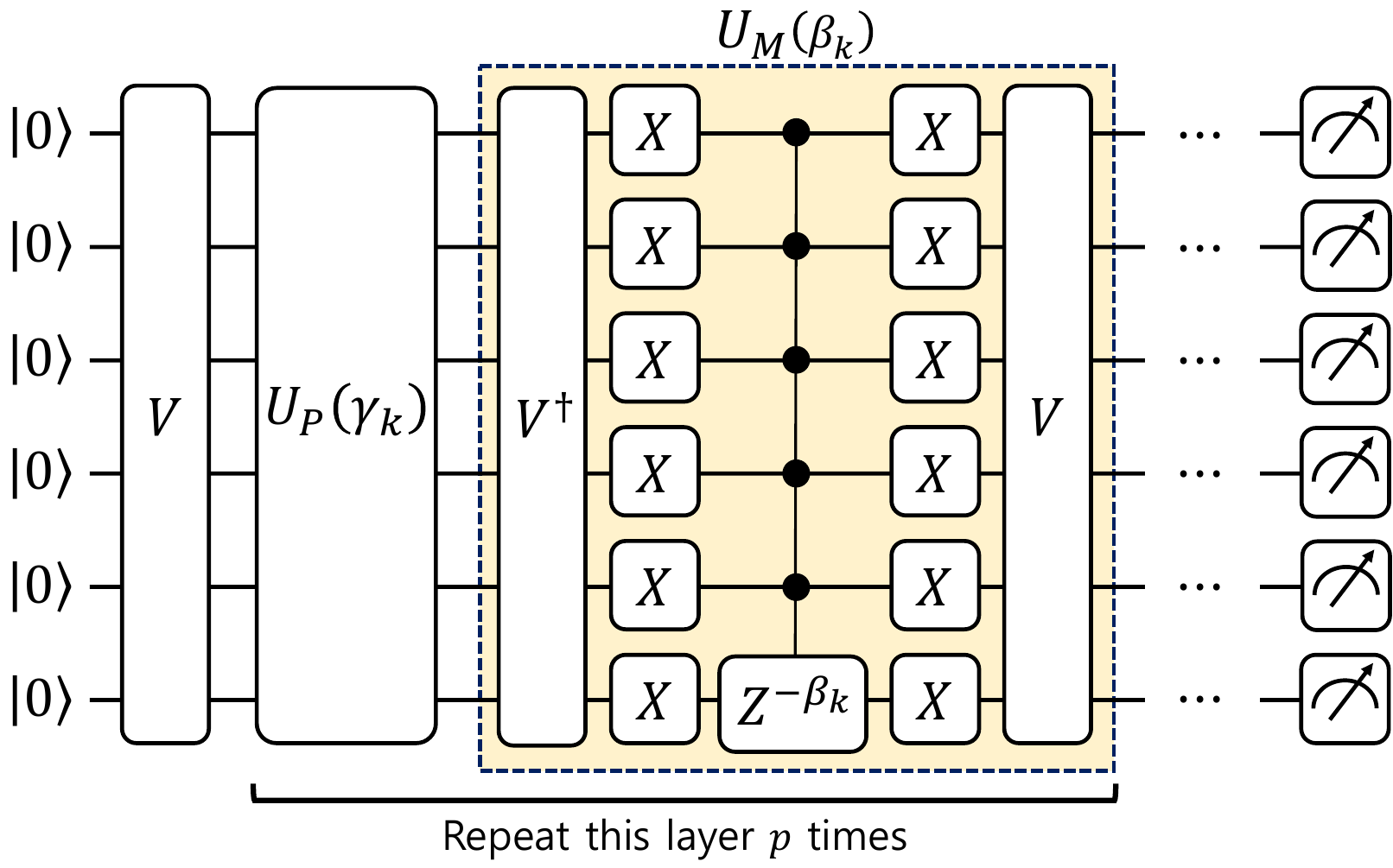}
\caption{
Schematic illustration of GM-QAOA. The circuit begins with the state-preparation unitary $V$, followed by alternating phase-separation unitaries $U_{P}(\gamma)$ and mixing unitaries $U_{M}(\beta)$.
Here, $X$ represents the Pauli-$X$ gate, and $Z^{\phi}$ is the phase-shift gate defined as $Z^{\phi}=\ket{0}\bra{0}+e^{i\phi}\ket{1}\bra{1}$.
The final state is measured in the computational basis to obtain approximate solutions.
}
\label{Figure01}
\end{figure}

Let $S \subseteq \{0, 1\}^{n}$ denote a subset of bitstrings over which the algorithm searches.
In many optimization problems, $S$ may correspond to the set of bitstrings satisfying some or all constraints of the problem.
The algorithm begins by preparing an initial state given by the uniform superposition over the bitstrings in $S$,
\begin{eqnarray}
\ket{\psi_{0}}&=&V\ket{0}^{\otimes n} \nonumber \\
&=&\frac{1}{\sqrt{|S|}}\sum_{x \in S} \ket{x},
\end{eqnarray}
where $V$ is a state-preparation unitary that admits an efficient quantum circuit implementation.
For a cost Hamiltonian $H_{C}$ encoding the objective function, the phase-separation unitary is defined as
\begin{equation}
U_{P}(\gamma)=e^{-i\gamma H_{C}}.
\end{equation}
In GM-QAOA, the mixer Hamiltonian is chosen as the rank-1 projector onto the initial state, 
\begin{equation}
\label{eq:rank1projector}
H_{M}=\ket{\psi_{0}}\bra{\psi_{0}},
\end{equation}
and the corresponding mixing unitary is
\begin{equation}
U_{M}(\beta)=e^{-i\beta H_{M}}.
\end{equation}
Starting from the initial state $\ket{\psi_{0}}$, the algorithm alternates between the phase-separation and mixing unitaries.
After $p$ layers, the variational state is given by
\begin{equation}
\ket{\psi_{p}(\bm{\beta}, \bm{\gamma})}=U_{M}(\beta_{p})U_{P}(\gamma_{p}) \cdots U_{M}(\beta_{1})U_{P}(\gamma_{1})\ket{\psi_{0}},
\end{equation}
where $\bm{\beta}=(\beta_{1}, \dots, \beta_{p})$ and $\bm{\gamma}=(\gamma_{1}, \dots, \gamma_{p})$ are variational parameters.
The overall structure of GM-QAOA is illustrated in Figure~\ref{Figure01}.
Finally, the state $\ket{\psi_{p}(\bm{\beta}, \bm{\gamma})}$ is measured in the computational basis, and the resulting bitstrings are used to obtain approximate solutions.

In GM-QAOA, both the phase-separation and mixing unitaries preserve the search space defined by the initial state.
Since the cost Hamiltonian $H_{C}$ is diagonal in the computational basis, the phase-separation unitary $U_{P}(\gamma)$ maps each basis state to itself up to a phase.
For the mixing unitary $U_{M}(\beta)$, since $H_{M}$ in Eq.~(\ref{eq:rank1projector}) is a rank-1 projector satisfying $H^{2}_{M}=H_{M}$, the matrix exponential can be written exactly as
\begin{equation}
\label{eq:exponent_init}
e^{-i\beta \ket{\psi_{0}}\bra{\psi_{0}}}=I-(1-e^{-i\beta})\ket{\psi_{0}}\bra{\psi_{0}},
\end{equation}
where $I$ denotes the identity operator.
From this expression, it is straightforward to see that $U_{M}(\beta)$ modifies only the amplitudes of the bitstrings in $S$, thereby preserving the search space defined by the initial state.
Consequently, both unitaries ensure that the overall computation remains confined to the search space.
We note that the mixing unitary $U_{M}(\beta)$ can be implemented as $V \cdot \bar{Z}_{0}^{-\beta} \cdot V^{\dagger}$, where $\bar{Z}_{0}^{-\beta}$ applies the phase $e^{-i \beta}$ to the state $\ket{0}^{\otimes n}$ and acts as the identity on all other computational basis states.
In practice, $\bar{Z}_{0}^{-\beta}$ is realized by sandwiching an $(n-1)$-controlled phase-shift gate $Z^{-\beta}$ between Pauli-$X$ gates applied to all qubits, as shown in Figure~\ref{Figure01}.

\subsection{A Product-Structured Variant of the Grover Mixer}
\label{subsec.II.2}
While the mixing unitary used in GM-QAOA preserves the search space defined by the initial state, its implementation requires a multi-controlled phase-shift gate, which incurs significant circuit overhead.
In particular, decomposing such a gate into elementary one- and two-qubit gates typically leads to deep quantum circuits, which can pose practical challenges for near-term quantum devices.
In this subsection, we consider initial states that admit a product structure over disjoint subsystems.
For such states, we propose a product-structured mixing unitary that preserves the search space while reducing the required circuit resources.

In many combinatorial optimization problems, the constraints specify a set of feasible bitstrings.
Within the context of GM-QAOA, these constraints can be incorporated into the initial state preparation to restrict the search space to feasible bitstrings.
However, enforcing all constraints simultaneously often leads to complex circuit constructions for the state-preparation unitary $V$.
A practical alternative is to encode only a subset of constraints into the initial state preparation, while the remaining constraints are handled via penalty terms in the cost Hamiltonian $H_{C}$.
This partial encoding can yield product-structured initial states, enabling significantly more efficient circuit implementations.
For instance, Bae \textit{et al.}~\cite{Bae2026} proposed a method for identifying independent qubit groups under linear constraints. 
Their method partitions the $n$ qubits into disjoint subsets such that each subset can be prepared independently.

\begin{figure}[t]
\centering
\includegraphics[width=0.47\textwidth]{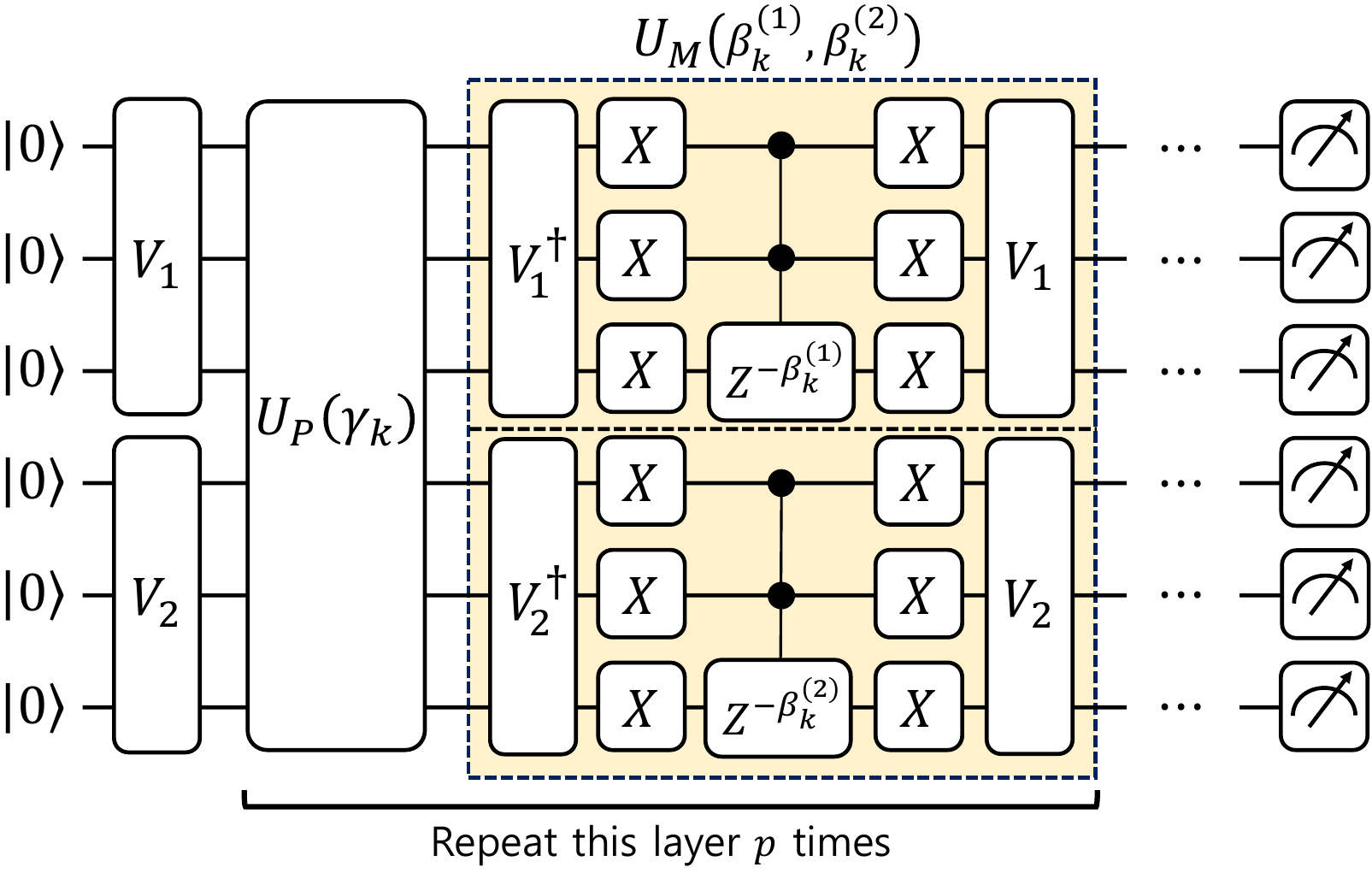}
\caption{
Circuit illustration of the product-structured variant of the Grover mixer for an initial state of product form $\ket{\psi_{0}}=\ket{\psi_{0}^{(1)}} \otimes \ket{\psi_{0}^{(2)}}$.
The state-preparation unitary decomposes as $V = V_{1} \otimes V_{2}$, where each $V_{j}$ independently prepares the $j$-th subsystem state $\ket{\psi_{0}^{(j)}}$.
The mixing unitary $U_{M}(\beta^{(1)}, \beta^{(2)})$ is implemented as two independent blocks, each realized using a multi-controlled phase-shift gate acting within its respective subsystem.
}
\label{Figure02}
\end{figure}

We now propose a product-structured mixing unitary tailored to initial states that factorize over disjoint qubit subsystems.
Suppose that the initial state can be written as
\begin{equation}
\ket{\psi_{0}}=\ket{\psi_{0}^{(1)}} \otimes \ket{\psi_{0}^{(2)}} \otimes \cdots \otimes \ket{\psi_{0}^{(\ell)}},
\end{equation}
where each state $\ket{\psi_{0}^{(j)}}$ is defined on a disjoint subset of qubits, and the state-preparation unitary decomposes accordingly as $V=V_{1} \otimes V_{2} \otimes \cdots \otimes V_{\ell}$.
Motivated by this product structure, we construct the mixing unitary as a tensor product of local mixing operations acting independently on each subsystem,
\begin{equation}
U_{M}(\beta^{(1)}, \dots, \beta^{(\ell)})=\bigotimes_{j=1}^{\ell} e^{-i\beta^{(j)}\ket{\psi^{(j)}_{0}}\bra{\psi^{(j)}_{0}}},
\end{equation}
where the parameters $\beta^{(j)}$ are independent across subsystems.
Each local unitary $e^{-i\beta^{(j)}\ket{\psi^{(j)}_{0}}\bra{\psi^{(j)}_{0}}}$ acts only on the corresponding subset of qubits and is implemented using a multi-controlled phase-shift gate within that subsystem, in the same manner as the global case, as illustrated in Figure~\ref{Figure02}.
This construction avoids the need for a multi-controlled operation acting on all qubits, replacing it with $\ell$ independent gates each acting on a smaller subset of qubits.
It follows from Eq.~(\ref{eq:exponent_init}) that each local unitary preserves the local subspace associated with $\ket{\psi^{(j)}_{0}}$.
Consequently, their tensor product $U_{M}$ preserves the product search space associated with $\ket{\psi_{0}}$.

In our approach, the variational state after $p$ layers is written as
\begin{equation}
\ket{\psi_{p}(\bm{\beta}^{(1)}, \dots, \bm{\beta}^{(\ell)}, \bm{\gamma})}=\left(\prod_{k=p}^{1}U_{M}(\beta_{k}^{(1)}, \dots, \beta_{k}^{(\ell)})U_{P}(\gamma_{k})\right)\ket{\psi_{0}},
\end{equation}
where $\bm{\beta}^{(j)}=(\beta_{1}^{(j)}, \dots, \beta_{p}^{(j)})$ and $\bm{\gamma}=(\gamma_{1}, \dots, \gamma_{p})$ denote the variational parameters.
The total number of variational parameters in a $p$-layer circuit is $p(\ell+1)$, compared to $2p$ in the original GM-QAOA.
While this represents an increase in the number of parameters, \rev{the local mixing construction requires fewer two-qubit gates than the global implementation after decomposition into elementary one- and two-qubit gates, which can facilitate implementation on near-term quantum devices.}
As demonstrated in Section~\ref{sec:results}, the proposed method can achieve convergence behavior comparable to that of the original GM-QAOA as a function of the number of layers $p$, suggesting that the reduced circuit complexity may be particularly advantageous in the presence of hardware noise.

%%%%%%%%%%%%%%%%%%%%%%%%%%%%%%%%%%%%%%%%%%%%%%%%%%%%%%%%%%%%%%%%%%%%
%%%
%%% Results and Discussion
%%%
%%%%%%%%%%%%%%%%%%%%%%%%%%%%%%%%%%%%%%%%%%%%%%%%%%%%%%%%%%%%%%%%%%%%

\section{Results and Discussion}
\label{sec:results}

\subsection{Simulation on the Exact-Cover Problem}
\label{subsec.III.1}

\begin{figure*}[t]
\centering
\includegraphics[width=1\textwidth]{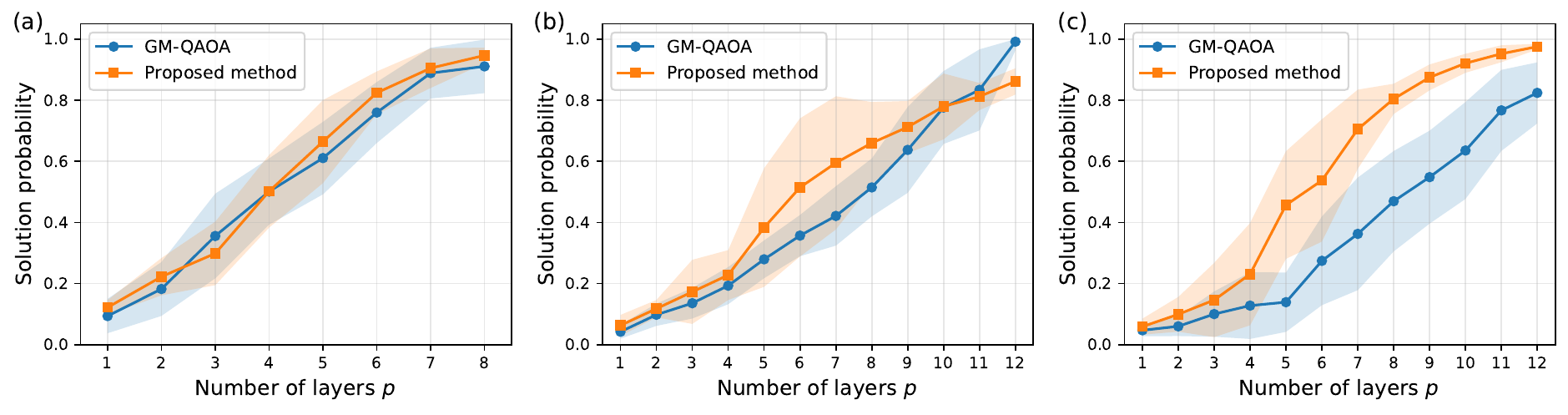}
\caption{
\rev{
Solution probability as a function of the number of layers $p$ for two exact-cover instances and a variant of the exact-cover formulation, using GM-QAOA and the proposed method.
Panels (a), (b), and (c) correspond to Examples 1, 2, and 3, respectively.
For each method and each value of $p$, the parameters are randomly initialized 30 times.
The markers indicate the mean solution probability over these runs, and the shaded bands represent the standard deviation.
}
}
\label{Figure03}
\end{figure*}

We first consider the exact-cover problem, which is a well-known constraint satisfaction problem.
Given a universe of $m$ elements and a collection of $n$ subsets, the exact-cover problem asks one to find a subcollection such that every element in the universe is covered exactly once.
This problem can be formulated as finding a binary vector $\bm{x} \in \{0, 1\}^{n}$ satisfying the linear constraint
\begin{equation}
A\bm{x}=\bm{1},
\end{equation}
where $A \in \{0, 1\}^{m \times n}$ is a binary matrix and $\bm{1}$ denotes the vector of length $m$ whose entries are all equal to $1$.
Here, each row of $A$ encodes which subsets contain a given universe element, so that $(A\bm{x})_{i}=1$ enforces that the $i$-th element is covered exactly once.
To apply QAOA, we reformulate the problem as a minimization.
When a feasible solution exists, satisfying the constraint $A\bm{x}=\bm{1}$ is equivalent to minimizing the quadratic objective
\begin{equation}
\label{eq:ecp_costftn}
\sum_{i=1}^{m}\left(\sum_{j=1}^{n}A_{i,j}x_{j} - 1\right)^{2},
\end{equation}
where $A_{i,j}$ denotes the $(i, j)$-th entry of $A$.
This formulation defines the cost Hamiltonian $H_{C}$.

\rev{As concrete examples, we first consider two exact-cover instances.
Example 1 is a seven-qubit instance defined by}
{\small
\renewcommand{\arraystretch}{0.7}
\begin{equation}
\label{eq:exact-cover instance_1}
\rev{A^{[1]}}=
\begin{pmatrix}
1&1&1&0&0&0&0\\
0&0&1&0&0&1&0\\
0&1&0&1&0&1&0\\
0&0&0&1&1&0&0\\
1&0&1&0&0&1&0\\
0&1&0&0&0&0&1\\
0&0&0&0&1&0&1
\end{pmatrix}.
\end{equation}
}
For this instance, the initial state can be prepared as
\begin{equation}
\label{eq:init_ECP_1}
\rev{\ket{\psi^{[1]}_{0}}}=\ket{W_{3}} \otimes \ket{W_{2}} \otimes \ket{++},
\end{equation}
using the method proposed by Bae \textit{et al.}~\cite{Bae2026}.
Here, \rev{$\ket{W_{\mu}}$ denotes the $\mu$-qubit W state,}
\begin{equation}
\ket{W_{\mu}}
= \frac{1}{\sqrt{\mu}}\sum_{\substack{s \in \{0, 1\}^{\mu} \\ \mathrm{wt}(s)=1}}\ket{s},
\end{equation}
where $\mathrm{wt}(s)$ denotes the Hamming weight of the binary string $s$,
and $\ket{+}=(\ket{0}+\ket{1})/\sqrt{2}$.
The state $\ket{W_{\mu}}$ can be prepared using a circuit with gate count and depth linear in $\mu$~\cite{Bartschi2019, Bae2026}.
\rev{
The two W-state registers enforce the constraints corresponding to the first and fourth rows of $A^{[1]}$, while the remaining constraints are included in the cost Hamiltonian.}

\rev{
Example 2 is a ten-qubit exact-cover instance defined by
{\small
\renewcommand{\arraystretch}{0.7}
\begin{equation}
\label{eq:exact-cover instance_2}
A^{[2]}=
\begin{pmatrix}
1&1&1&1&0&0&0&0&0&0\\
0&0&1&0&0&0&1&1&0&0\\
0&1&0&0&1&0&0&0&0&1\\
0&0&0&0&1&1&1&0&0&0\\
1&0&1&0&0&0&1&0&0&1\\
0&1&0&0&0&1&0&0&1&0\\
0&0&0&1&0&0&0&1&1&1
\end{pmatrix}.
\end{equation}
}
For this instance, we prepare the initial state
\begin{equation}
\label{eq:init_ECP_2}
\ket{\psi^{[2]}_{0}}=\ket{W_{4}} \otimes \ket{W_{3}} \otimes \frac{1}{2}(\ket{000}+\ket{001}+\ket{010}+\ket{100}),
\end{equation}
where the first two registers enforce the constraints corresponding to the first and fourth rows of $A^{[2]}$.
The last register restricts $x_{8}+x_{9}+x_{10} \le 1$, a necessary condition implied by the constraint corresponding to the seventh row of $A^{[2]}$.
Since this restriction does not enforce that constraint in full, all constraints except those corresponding to the first and fourth rows remain in the cost Hamiltonian.
}

\rev{
Example 3 considers a variant of the exact-cover formulation with integer-valued coefficients and a modified right-hand side,
\begin{equation}
A^{[3]}\bm{x}=4\cdot\bm{1},\quad \bm{x} \in \{0, 1\}^{8},
\end{equation}
with
{\small
\renewcommand{\arraystretch}{0.7}
\begin{equation}
\label{eq:exact-cover instance_3}
\rev{A^{[3]}}=
\begin{pmatrix}
1&1&3&2&0&0&2&0\\
0&2&2&3&1&1&0&0\\
2&0&2&0&0&0&1&1\\
0&0&2&0&0&2&1&1\\
0&0&2&0&2&2&0&0
\end{pmatrix}.
\end{equation}
}
The cost function is defined analogously to Eq.~\eqref{eq:ecp_costftn}, with the target value 1 replaced by 4 in each squared term.
Reducing the first three constraints modulo two yields even-parity conditions on the disjoint registers.
We therefore prepare the product state
\begin{equation}
\label{eq:init_ECP_3}
\ket{\psi^{[3]}_{0}}=\ket{GHZ_{3,0}^{(X)}} \otimes \ket{GHZ_{3,0}^{(X)}} \otimes \ket{GHZ_{2,0}^{(X)}},
\end{equation}
where
\begin{eqnarray}
\label{eq:GHZ}
\ket{GHZ_{\mu,\nu}^{(X)}}
&=&\frac{1}{\sqrt{2}}\left(\ket{+}^{\otimes \mu} + (-1)^{\nu}\ket{-}^{\otimes \mu}\right) \nonumber \\
&=&\frac{1}{\sqrt{2^{\mu-1}}}\sum_{\substack{s \in \{0, 1\}^{\mu} \\ \mathrm{wt}(s)\equiv \nu \pmod{2}}}\ket{s},
\end{eqnarray}
with $\nu \in \{0, 1\}$ and $\ket{-}=(\ket{0}-\ket{1})/\sqrt{2}$.
Since the initial state encodes only parity conditions, all five constraints are retained in the cost function.
In all three examples, the initial state has a product structure over three disjoint subsets of qubits.
Accordingly, the proposed mixing unitary consists of three local operations, each acting independently on one subset. 
This results in $p$ phase parameters and $3p$ mixing parameters for a total of $4p$ variational parameters in a $p$-layer circuit.
}

\rev{
The simulations are carried out in Python 3.13.15 using PennyLane 0.45.1~\cite{Bergholm2018}, and parameter optimization is performed using the L-BFGS-B optimizer~\cite{Liu1989, Byrd1995} implemented in SciPy 1.18.1.
All simulations are noiseless, and expectation values are evaluated without finite-shot sampling.
For a fair comparison between the original GM-QAOA and the proposed method, the same initial state and cost Hamiltonian are used for both methods within each instance, so that they differ only in the structure of the mixing unitary.
For each method and each value of $p$, the parameters are randomly initialized 30 times, with each parameter sampled uniformly from $[0, 2\pi)$.
No bounds are imposed on the variational parameters during optimization.
Gradients are estimated by forward finite differences with a step size of $10^{-8}$.
The stopping tolerances are set to $\text{ftol}=2.22\times10^{-9}$ and $\text{gtol}=10^{-5}$, with maximum limits of $1.5 \times 10^{4}$ iterations and $4 \times 10^{5}$ function evaluations per run to allow convergence under these criteria.
The resulting solution probabilities are shown in Figure~\ref{Figure03}.
}

\rev{The proposed method exhibits broadly comparable convergence to that of the original GM-QAOA as the number of QAOA layers increases for Examples 1 and 2, as shown in Figures~\ref{Figure03}(a) and \ref{Figure03}(b).
For Example 3, the proposed method achieves higher mean solution probabilities over the range of $p$ considered in these simulations, as shown in Figure~\ref{Figure03}(c).}
Although the proposed ansatz introduces more variational parameters, its circuit implementation requires fewer entangling gates when decomposed into elementary one- and two-qubit gates.
\rev{Circuit resources are evaluated using PennyLane's compilation procedure with the gate set $\{RX, RY, RZ, \mathrm{CNOT}\}$, applying rotation merging and inverse-gate cancellation over two passes.
No ancillary qubits are used, and arbitrary two-qubit connectivity is assumed.
Under these compilation settings, the number of $\mathrm{CNOT}$ gates required to implement the mixing unitary is reduced from 218 to 28, from 818 to 74, and from 354 to 24 for Examples 1, 2, and 3, respectively.
These reductions of approximately 87\%, 91\%, and 93\% result from replacing the global multi-controlled phase-shift gate with local operations on each subsystem.}

On the other hand, the increased number of parameters leads to a larger number of circuit executions during the optimization process.
\rev{
At the largest value of $p$ considered for each instance, the mean numbers of function evaluations for GM-QAOA and the proposed method are approximately $4.64 \times 10^{3}$ and $7.84 \times 10^{3}$ for Example 1, $1.22 \times 10^{4}$ and $2.05 \times 10^{4}$ for Example 2, and $9.50 \times 10^{4}$ and $1.23 \times 10^{5}$ for Example 3, respectively.
These counts include evaluations used to estimate gradients by finite differences.
Nevertheless, at these values of $p$, the reduction in CNOT gates per mixing unitary more than offsets the increased number of function evaluations, yielding a lower estimated cumulative CNOT count for the mixing operations over the optimization process.
Together with the reduced gate count per circuit, this suggests that the proposed method may be more favorable for near-term quantum devices where gate noise is a primary limiting factor.
}

\rev{
For blocks initialized in a W state, the local Grover mixer is equivalent to a complete-graph XY mixer within the single-excitation subspace, up to a rescaling of the mixing parameter and a global phase.
To see this, let $\ket{e_{j}}$ denote the computational basis state with a single excitation at qubit $j$.
The complete-graph XY Hamiltonian is defined as~\cite{Wang2020}
\begin{equation}
H_{XY}=\frac{1}{2}\sum_{i<j}(X_{i}X_{j}+Y_{i}Y_{j}),
\end{equation}
where $X_{i}$ and $Y_{i}$ are Pauli operators acting on qubit $i$.
Each term exchanges an excitation between two qubits, preserving the single-excitation subspace $\mathcal{H}_{1}=\operatorname{span}\{\ket{e_j}\}_{j=1}^{\mu}$.
Since every pair of qubits is coupled, its restriction to this subspace satisfies
\begin{equation}
H_{XY}|_{\mathcal{H}_{1}}=\sum_{i \neq j}\ket{e_{i}}\bra{e_{j}}=\mu\ket{W_{\mu}}\bra{W_{\mu}}-I_{\mathcal{H}_{1}}.
\end{equation}
Consequently, within $\mathcal{H}_{1}$,
\begin{equation}
e^{-i\theta H_{XY}}=e^{i\theta}e^{-i\mu\theta\ket{W_{\mu}}\bra{W_{\mu}}}.
\end{equation}
Thus, setting $\beta=\mu\theta$ reproduces the local Grover mixer, apart from an irrelevant global phase.
Therefore, for W-state blocks, the local Grover mixer can be implemented using the equivalent complete-graph XY evolution when this offers a lower circuit cost for the native gate set and connectivity of the target device.
}

\rev{
However, while standard XY mixers preserve Hamming weight, the Grover mixer construction can mix computational basis states with different Hamming weights within a prescribed search space.
Given a circuit preparing a uniform superposition over the chosen basis states, the corresponding Grover mixer preserves their span.
This allows mixing between Hamming weights zero and one in the last block of Example 2, as well as between different even Hamming weights within the blocks of Example 3.
Such transitions cannot be generated by a standard XY mixer.
Different mixer implementations may therefore be used for different blocks while retaining the product structure of the proposed ansatz.
}

\rev{
The LX-mixer framework introduced by Fuchs and Bassa~\cite{Fuchs2024} generalizes X and XY mixers and provides a relaxation of the Grover mixer, offering a systematic approach to constructing resource-efficient mixers that preserve a specified feasible subspace.
In the present construction, each local Grover unitary is implemented using the corresponding subsystem state-preparation circuit, its inverse, and a phase shift on the all-zero state of that subsystem.
This provides a direct implementation when such preparation circuits are available, while allowing us to investigate how independent subsystem parameters affect solution quality and optimization overhead.
Within the product structure considered here, suitably constructed LX mixers could replace individual local Grover mixers while retaining independent variational parameters for each subsystem.
This provides a possible extension of the proposed framework, with potential circuit-resource savings depending on the structure of the local feasible subspaces and the chosen implementation.
The resulting optimization performance would need to be investigated separately.
}

\subsection{Performance Evaluation on the TSP}
\label{subsec.III.2}

\begin{figure*}[t]
\centering
\includegraphics[width=1\textwidth]{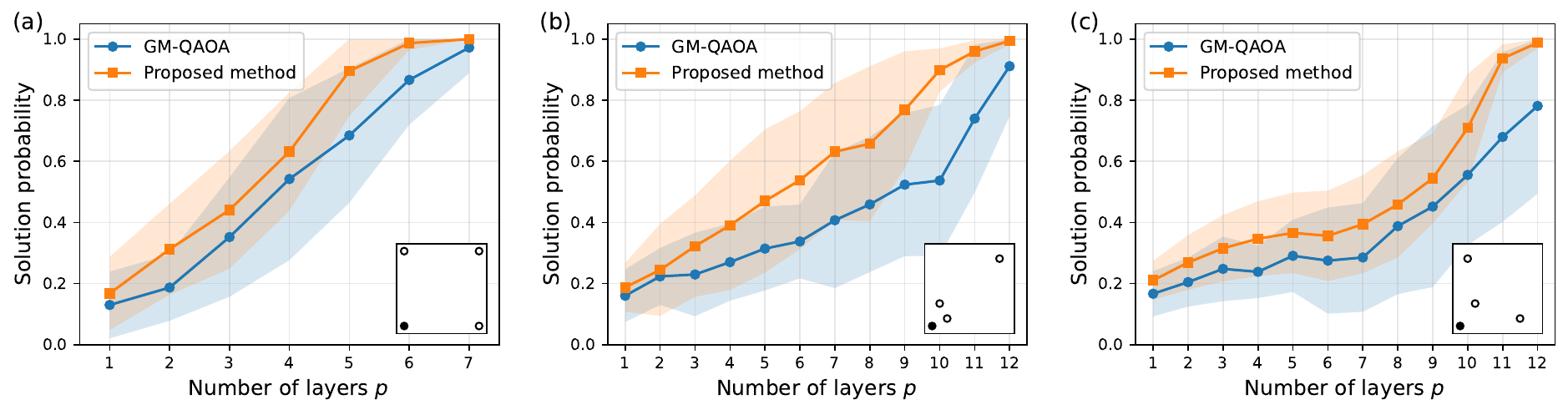}
\caption{
Solution probability as a function of the number of layers $p$ for the TSP using GM-QAOA and the proposed method.
Plots (a), (b), and (c) correspond to three instances of the four-city TSP (nine qubits) with different city locations, as illustrated in the insets, where circles represent cities and the filled circle indicates the starting and ending point of the tour.
The simulation setting follows that of Figure~\ref{Figure03}.
Across all three instances, 
\rev{
the proposed method achieves broadly comparable or higher mean solution probabilities than GM-QAOA over the range of $p$ considered.
}
}
\label{Figure04}
\end{figure*}

The traveling salesman problem~(TSP) is a combinatorial optimization problem that seeks a minimum-cost tour visiting each city exactly once and returning to the starting point.
We consider the following formulation for an instance with $n+1$ cities.
Let $x_{t,i}$ be a binary variable such that $x_{t,i}=1$ if the $i$-th city is visited at time step $t$, and $x_{t,i}=0$ otherwise, where $0 \le t, i \le n$. 
The starting city is fixed as $i=0$, that is, $x_{0, 0}=1$ and $x_{0, i}=0$ for $i \neq 0$.
Under this formulation, a valid tour corresponds to a permutation of cities across the $n$ time steps.
Assuming that the travel cost $\phi_{i,j}$ from the $i$-th city to the $j$-th city is symmetric, that is, $\phi_{i,j}=\phi_{j,i}$ for $0 \le i, j \le n$, the travel cost of a tour can be expressed as
\begin{equation}
T(\bm{x}) = \sum_{i=1}^{n}\phi_{0,i}x_{1,i} + \sum_{t=1}^{n-1}\sum_{i,j=1,i\neq j}^{n}\phi_{i,j}x_{t,i}x_{t+1,j} + \sum_{i=1}^{n}\phi_{i,0}x_{n,i}.
\end{equation}
The feasible subspace is defined by two sets of constraints.
First, exactly one city must be visited at each time step,
\begin{equation}
P_{1}(\bm{x})=\sum_{t=1}^{n}\left(\sum_{i=1}^{n}x_{t,i}-1\right)^{2}=0,
\end{equation}
and second, each city must appear exactly once in the tour,
\begin{equation}
P_{2}(\bm{x})=\sum_{i=1}^{n}\left(\sum_{t=1}^{n}x_{t,i}-1\right)^{2}=0.
\end{equation}

The above formulation requires a total of $n^{2}$ qubits, and the search space has size $2^{n^{2}}$ when the standard initial state $\ket{+}^{\otimes n^{2}}$ is used, in which case both constraints $P_{1}$ and $P_{2}$ must be enforced through penalty terms in the cost function.
Instead, we encode the time-step constraint $P_{1}(\bm{x})=0$ into the initial state preparation.
After relabeling the variables according to the time-step index as $x_{n(t-1)+i}$, the $n^{2}$ variables are divided into $n$ blocks, each corresponding to one time step.
Since $P_{1}(\bm{x})=0$ requires exactly one city to be visited in each block, each block can be prepared in the state $\ket{W_{n}}$, yielding the product form
\begin{equation}
\ket{W_{n}}\otimes \ket{W_{n}} \otimes \cdots \otimes \ket{W_{n}},
\end{equation}
which reduces the effective search space to $n^{n}$.
Since the constraint $P_{2}$ is not incorporated into the initial state, it is enforced through a penalty term in the cost function,
\begin{equation}
\tilde{T}(\bm{x})=T(\bm{x})+\lambda P_{2}(\bm{x}),
\end{equation}
where $\lambda>0$ is a penalty coefficient, which we set to twice the maximum pairwise travel cost, $\lambda=2\max_{i,j}\phi_{i,j}$.
Due to the product structure of the initial state, the mixing unitary in the proposed method can be implemented using $n$ multi-controlled phase-shift gates, each acting on a subsystem.
In this case, the proposed QAOA circuit introduces a total of $p(n+1)$ variational parameters for $p$ layers.
One may also consider incorporating both constraints into the initial state preparation to further reduce the search space.
However, such an approach no longer admits a product structure, and the proposed method cannot be applied. 
As will be discussed later, the proposed approach with partial constraint encoding may be more favorable than incorporating both constraints into the initial state, from the perspective of circuit complexity.

\subsubsection{Convergence Behavior and Resource Trade-offs}

\begin{figure*}[t]
\centering
\includegraphics[width=1\textwidth]{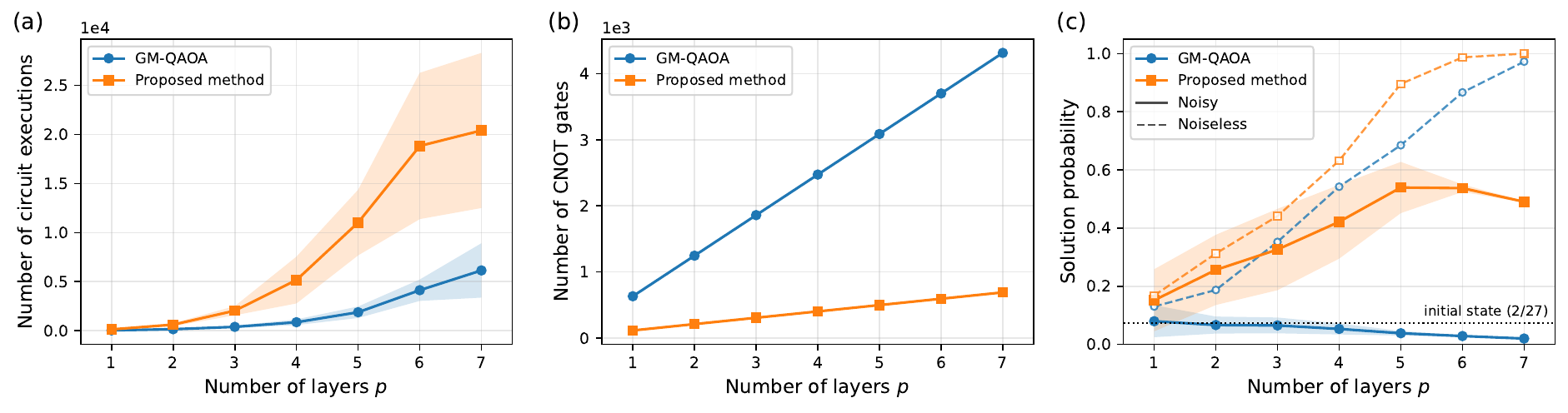}
\caption{
Resource trade-offs \rev{and noise robustness of} GM-QAOA and the proposed method for \rev{the four-city TSP instance shown in Figure~\ref{Figure04}(a)}, as a function of the number of layers $p$.
(a) Total number of circuit executions required during optimization.
(b) Number of $\mathrm{CNOT}$ gates required \rev{in the full circuit}, obtained after decomposition into the $\{RX, RY, RZ, \mathrm{CNOT}\}$ gate set.
(c) \rev{Solution probability evaluated under a depolarizing noise model with one- and two-qubit gate error rates $\epsilon_{1}=10^{-4}$ and $\epsilon_{2}=10^{-3}$. 
The parameters obtained from noiseless optimization are kept fixed, and noise is applied after every gate of the decomposed circuit.
Solid and dashed curves indicate noisy and noiseless results, respectively.
The dotted horizontal line in (c) marks the noiseless initial-state solution probability.
In (a) and (c), markers indicate the mean over 30 random initializations, and shaded bands represent the standard deviation in (a) and for the noisy results in (c).}
}
\label{Figure05}
\end{figure*}

We evaluate the proposed method on three instances of the four-city TSP, which require nine qubits under the above formulation.
Each instance corresponds to a different set of city locations, as illustrated in the insets of Figure~\ref{Figure04}.
The simulation settings are the same as in Section~\ref{subsec.III.1}.
\rev{As shown in Figure~\ref{Figure04}, the proposed method achieves broadly comparable or higher mean solution probabilities than GM-QAOA across all three instances.
Together with the results for the exact-cover problem, this suggests that the proposed method can maintain competitive solution quality across the problem types and instances considered here.}

A rigorous theoretical understanding of \rev{this performance} despite the relaxed mixing structure remains an open question. 
One possible explanation is that the additional degrees of freedom introduced by the independent mixing parameters may allow more flexible amplitude modulation within each subsystem, potentially compensating for the relaxed global mixing structure.
% We also note that the shaded bands in Figure~\ref{Figure04} tend to be somewhat wider for the proposed method in some instances, which may reflect a more complex optimization landscape arising from the increased number of parameters.
In the proposed framework, however, the granularity of the product structure can in principle be adjusted.
For example, the initial state $\ket{W_{3}} \otimes \ket{W_{3}} \otimes \ket{W_{3}}$ considered here has $\ell=3$ subsystems, but one could alternatively treat the first two subsystems as a single larger subsystem, yielding $\ell=2$ with two multi-controlled phase-shift gates.
This flexibility allows the number of subsystems to be chosen so as to balance circuit complexity against the optimization overhead introduced by the additional variational parameters.

We further analyze the resource trade-offs between the two methods for \rev{the TSP instance in Figure~\ref{Figure04}(a)}, as shown in Figure~\ref{Figure05}.
Figure~\ref{Figure05}(a) shows the total number of circuit executions required during optimization as a function of $p$.
A circuit execution corresponds to one evaluation of a quantum circuit, either for estimating the cost function or gradients.
The accumulated number of such evaluations until convergence is used as a measure of optimization overhead.
Figure~\ref{Figure05}(b) presents the number of $\mathrm{CNOT}$ gates required \rev{in the full circuit, including initial state preparation, cost unitaries, and mixing unitaries,} obtained after decomposition into the $\{RX, RY, RZ, \mathrm{CNOT}\}$ gate set.
Figure~\ref{Figure05}(c) reports the \rev{solution probability under} a depolarizing noise model with one- and two-qubit gate error rates $\epsilon_{1}=10^{-4}$ and $\epsilon_{2}=10^{-3}$, respectively.
\rev{Using the parameters obtained from noiseless optimization, we evaluate the full circuit with depolarizing noise applied after each decomposed gate.
Solution probabilities are evaluated directly from the simulated density matrix without finite-shot sampling.
Gate errors during state preparation are included, while reset and readout errors are not modeled.
The results are averaged over 30 independent optimization runs with random initializations, and the corresponding noiseless solution probabilities are shown for comparison.
This comparison provides insight into how gate noise affects the solution probabilities of the two methods.}

As shown in Figure~\ref{Figure05}(a), the proposed method requires substantially more circuit executions than GM-QAOA, and the gap \rev{generally} widens as $p$ increases.
This is \rev{partly due to} the larger number of variational parameters introduced by the proposed method.
Figure~\ref{Figure05}(b) likewise reflects the reduced circuit complexity of the proposed mixing structure.
Replacing the global multi-controlled phase-shift gate with local operations on each subsystem reduces the number of CNOT gates from 572 to 54 per mixing operator.
\rev{The number of single-qubit gates per mixing operator is also reduced from 1,121 to 87.}
While each of these observations follows naturally from the construction, their significance becomes clear in light of the comparable convergence behavior.
\rev{While achieving} a comparable level of optimization performance, the proposed method demands more circuit executions but operates with considerably lower circuit complexity and \rev{reduced exposure to gate noise}, whereas GM-QAOA requires fewer circuit executions at the cost of deeper and more noise-prone circuits.
\rev{Combining the mean number of circuit evaluations with the CNOT count of the full circuit yields a lower estimated cumulative CNOT cost for the proposed method throughout the range of $p$ considered.
For example, at $p=7$, this estimate is approximately $1.41 \times 10^{7}$, compared with $2.65 \times 10^{7}$ for GM-QAOA.}
As illustrated in Figure~\ref{Figure05}(c), 
\rev{
the proposed method achieves substantially higher mean solution probabilities than GM-QAOA under the depolarizing noise model throughout the range of $p$ considered.
These results suggest that the reduced circuit complexity can help mitigate the impact of gate noise, making the proposed method a potentially favorable option for near-term quantum devices.
}

\begin{figure}[t]
\centering
\includegraphics[width=0.47\textwidth]{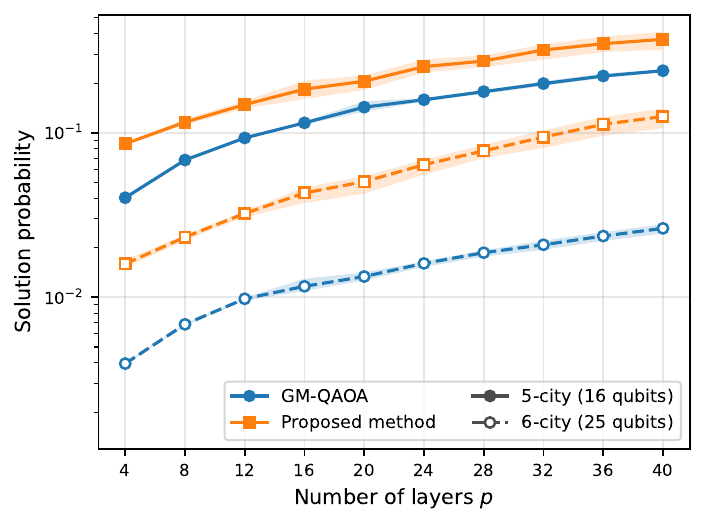}
\caption{
\rev{
Solution probability as a function of the number of layers $p$ for GM-QAOA and the proposed method using linear-ramp parameterizations on five- and six-city TSP instances, requiring 16 and 25 qubits, respectively.
For each method, instance, and value of $p$, the ramp coefficients are selected by minimizing the cost function over the parameter grid.
Solid curves with filled markers and dashed curves with open markers correspond to the five- and six-city cases, respectively.
Markers indicate the mean solution probability over three instances with different city locations, and shaded bands span the minimum and maximum values across these instances.
All results are obtained under noiseless conditions, and the vertical axis uses a logarithmic scale.
}
}
\label{Figure06}
\end{figure}

\rev{
To further examine the performance of the proposed method on larger instances, we consider three instances each of the five- and six-city TSP, requiring 16 and 25 qubits, respectively.
The city coordinates are provided in the accompanying GitHub repository.
We employ a linear-ramp parameterization adapted from the approach of Monta{\~n}ez-Barrera and Michielsen~\cite{Montanez2025}.
For a $p$-layer circuit, we define $t_{k}=(k - 1/2)/p$ for $k=1, \dots, p$, and set $\gamma_{k}=at_{k}$ and $\beta_{k}=b(1-t_{k})$ for GM-QAOA.
The proposed method uses the same cost parameterization and independent mixing coefficients for each subsystem, $\beta_{k}^{(j)}=b_{j}(1-t_{k})$.
To assess the solution quality attainable within each linear-ramp parameterization, we perform an exhaustive grid search over the corresponding ramp coefficients, using 20 points per axis, $a \in \{0.1, 0.2, \dots, 2.0\}$ and $b,b_{j} \in [-2\pi, 0]$.
Among the grid points, we select the one that minimizes the cost expectation value and report the corresponding noiseless solution probability.
}

\rev{
Figure~\ref{Figure06} presents the mean solution probabilities over the three instances for each problem size.
Both methods exhibit increasing mean solution probabilities over the range of $p$ considered, with the proposed method achieving higher mean values for both the five- and six-city cases.
These results show that competitive solution probabilities remain accessible within the restricted linear-ramp parameterization on larger instances, suggesting that the competitive performance observed for the proposed method may extend beyond the four-city examples.
}

\subsubsection{Comparison Between Partial and Full Constraint Encodings}

\begin{figure*}[t]
\centering
\includegraphics[width=0.8\textwidth]{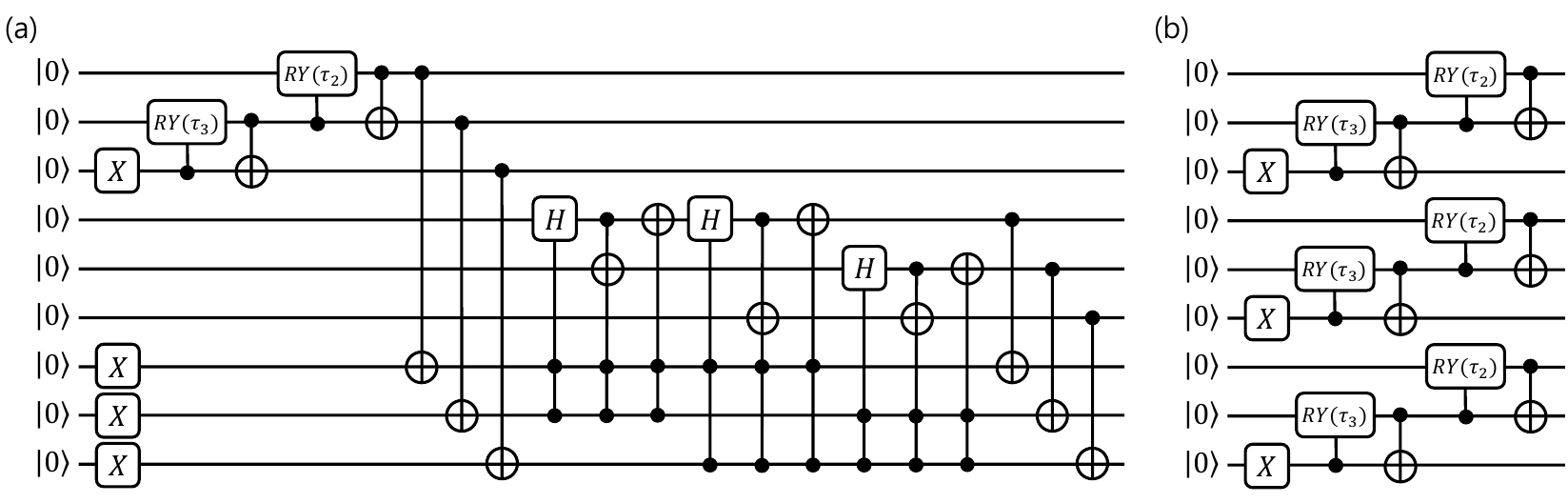}
\caption{
State preparation circuits for the four-city TSP instance (nine qubits) with (a) full constraint encoding ($P_{1}$ and $P_{2}$) and (b) partial constraint encoding ($P_{1}$).
In (a), the circuit prepares an equal superposition over all $3!=6$ valid tours.
In (b), three locally acting blocks each prepare a $\ket{W_{3}}$ state on a disjoint subset of qubits, spanning a search space of $3^{3}=27$ states.
Here, $\tau_{s}=2\arccos(\sqrt{1/s})$.
}
\label{Figure07}
\end{figure*}

One might expect that incorporating both constraints $P_{1}$ and $P_{2}$ into the initial state preparation is preferable, as this reduces the effective search space from $n^{n}$ to $n!$.
Reducing the search space in this way can lower the number of layers $p$ required to achieve a high solution probability.
However, preparing such an initial state generally increases the complexity of the state preparation circuit.
Furthermore, since the resulting initial state no longer admits a product structure, the proposed product-structured mixing unitary cannot be applied.
Here we compare the circuit complexity of the two encodings at the values of $p$ where they achieve high solution probabilities.
As shown below, incorporating full constraints ($P_{1}$ and $P_{2}$) may be less efficient than partial constraint encoding ($P_{1}$).

Figure~\ref{Figure07} illustrates the state preparation circuits for both encodings on the four-city TSP instance.
The full constraint encoding (Figure~\ref{Figure07}(a)) requires preparing an equal superposition over all $n!$ valid tours, and we adopt the explicit construction proposed by B\"{a}rtschi and Eidenbenz~\cite{Bartschi2020} for this purpose.
The partial encoding (Figure~\ref{Figure07}(b)) decomposes into $n$ independent blocks, each preparing a $\ket{W_{n}}$ state on a disjoint subset of qubits, resulting in a substantially simpler circuit.
The two encodings also differ in the complexity of their phase-separation and mixing unitaries.
In the full constraint encoding, the phase-separation unitary involves only the bare objective $T(\bm{x})$ without a penalty term, making it relatively simple, but the mixing unitary requires a global multi-controlled phase-shift gate acting on all nine qubits.
In the partial encoding, the penalty term in $\tilde{T}(\bm{x})$ adds complexity to the phase-separation unitary, while the mixing unitary decomposes into local operations on each subsystem.

\begin{table}[t]
\centering
\begin{tabular}{lcc}
\hline
\hline
& Full ($P_{1}$ \& $P_{2}$) & Partial ($P_{1}$) \\
\hline
Optimal $p$ & 1 & 7 \\
Total one-qubit gates & \rev{2077} & 834 \\
Total two-qubit gates & \rev{1010} & 690 \\
Circuit depth & \rev{2084} & 450 \\
\hline
\hline
\end{tabular}
\caption{
Comparison of circuit complexity between full and partial constraint encoding for TSP instance (a), evaluated at the optimal $p$, taken as the smallest value of $p$ at which each method achieves a mean solution probability exceeding 0.99.
All results are obtained after decomposing the entire circuit into the $\{RX, RY, RZ, \mathrm{CNOT}\}$ gate set.
}
\label{Table01}
\end{table}

For the TSP instance in Figure~\ref{Figure04}(a), the full constraint encoding achieves a high solution probability (mean solution probability exceeding 0.99) at $p=1$, whereas the partial encoding requires $p=7$ layers to reach a comparable solution probability.
Despite requiring fewer layers, the full encoding incurs substantially higher circuit complexity, as summarized in Table~\ref{Table01}.
The gate counts here reflect the entire circuit, including state preparation, phase-separation, and mixing unitaries, after decomposition into the $\{RX, RY, RZ, \mathrm{CNOT}\}$ gate set.
At their respective values of $p$, the full encoding requires \rev{2077} one-qubit gates and \rev{1010} two-qubit gates, compared to 834 one-qubit gates and 690 two-qubit gates for the partial encoding.
The difference is even more pronounced in circuit depth, with the full encoding reaching a depth of \rev{2084}, about \rev{4.6} times larger than the depth of 450 for the partial encoding.
These results indicate that, \rev{for this instance,} the reduction in the number of layers does not compensate for the increased circuit complexity, and that partial constraint encoding achieves a more favorable overall circuit complexity.
\rev{
For larger TSP instances, the state-preparation circuit for the full constraint encoding has $O(n^{2})$ depth and requires $O(n^{3})$ gates~\cite{Bartschi2020}, whereas the partial constraint encoding has $O(n)$ depth and requires $O(n^{2})$ gates when the $n$ independent registers are prepared in parallel.
Since the state-preparation circuit and its inverse also appear in the mixing unitaries, this difference in scaling may contribute to lower overall circuit costs for larger instances.
}

\subsubsection{Comparison with Alternative Mixers}

\begin{figure*}[t]
\centering
\includegraphics[width=1\textwidth]{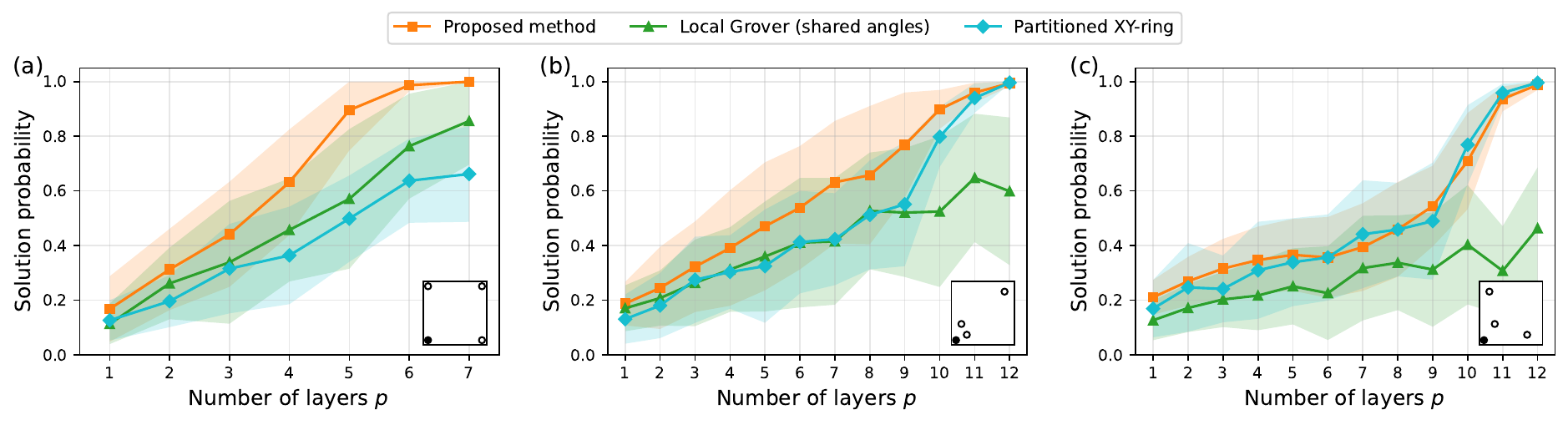}
\caption{
\rev{
Comparison of the proposed method, the shared-angle local Grover configuration, and the partitioned XY-ring mixer for the four-city instances shown in Figure~\ref{Figure04}.
The shared-angle local Grover configuration uses one mixing angle per layer across all subsystems, whereas the proposed method and the partitioned XY-ring mixer use independent mixing angles for each subsystem.
The simulation settings are the same as those used for Figure~\ref{Figure04}, and the partitioned XY-ring mixer uses the same initial parameter values as the proposed method for each corresponding run.
Markers indicate the mean over 30 random initializations, and shaded bands represent the standard deviation.
}
}
\label{Figure08}
\end{figure*}

\rev{
We compare the proposed method with two alternative mixer configurations on the four-city TSP instances considered above.
The first is a shared-angle local Grover configuration, following the product construction considered by Fuchs \textit{et al.}~\cite{Fuchs2025}.
It uses the same local Grover mixing operators as the proposed method but shares a single mixing angle across all subsystems within each layer, reducing the number of variational parameters from $4p$ to $2p$.
The second uses a partitioned XY-ring mixer~\cite{Wang2020}, in which the two-qubit operations $\exp[-i\beta_{k}^{(r)}(X_{i}X_{j}+Y_{i}Y_{j})/2]$ are applied sequentially along the edges $(0,1)$, $(1,2)$, and $(2,0)$ of each three-qubit subsystem $r$, giving the same number of variational parameters as the proposed method.
Although the three-qubit ring has the same edges as a complete graph, this ordered product generally differs from the simultaneous XY evolution discussed in Sec.~\ref{subsec.III.1}.
The simulation settings are the same as those used for Figure~\ref{Figure04}, and the partitioned XY-ring mixer uses the same initial parameter values as the proposed method for each corresponding run.
}

\rev{
Figure~\ref{Figure08} shows that the proposed method achieves higher mean solution probabilities than the shared-angle local Grover configuration across all three instances throughout the range of $p$ considered.
However, the shared-angle configuration requires fewer circuit evaluations during optimization and can be advantageous when a lower solution probability is acceptable.
For instance, in Figure~\ref{Figure08}(a), it achieves a mean solution probability of approximately 0.856 at $p=7$ with 2,214 circuit evaluations on average, whereas the proposed method reaches approximately 0.999 with 20,405 evaluations.
Thus, the additional optimization cost should be considered together with the improvement in solution probability.
Comparisons at similar solution probabilities reveal a further trade-off between optimization overhead and the resources required per circuit.
In Figure~\ref{Figure08}(b), the proposed method achieves a mean solution probability of 0.632 at $p=7$, compared with 0.599 for the shared-angle configuration at $p=12$.
Although the proposed method requires more circuit evaluations on average, approximately 20,460 versus 9,442, its full circuit contains fewer CNOT gates, 690 versus 1,170.
The proposed method may therefore be preferable when reducing circuit complexity and exposure to gate noise per evaluation is a priority, while parameter sharing may be preferable when minimizing optimization overhead is more important.
}

\rev{
The comparison with the partitioned XY-ring mixer exhibits a stronger dependence on the problem instance.
In Figure~\ref{Figure08}(a), the proposed method achieves a mean solution probability close to unity at $p=7$, whereas the partitioned XY-ring mixer reaches approximately 0.662.
In Figures~\ref{Figure08}(b) and \ref{Figure08}(c), however, both methods achieve mean solution probabilities above 0.98 at $p=12$.
Thus, the advantage observed for the proposed method in Figure~\ref{Figure08}(a) does not extend uniformly to the other instances.
The resource comparison further illustrates this dependence.
Under the same compilation procedure, the partitioned XY-ring circuit at $p=12$ requires 954 CNOT gates, compared with 1,170 for the proposed method, although it uses more single-qubit gates, 1,743 versus 1,419.
The partitioned XY-ring mixer also requires approximately one-third as many circuit evaluations during optimization at this value of $p$.
Consequently, the partitioned XY-ring mixer can offer an effective alternative for some instances, combining high solution probability with lower CNOT counts and optimization overhead.
}

\rev{
These comparisons show that alternative mixer configurations can offer favorable resource trade-offs, with comparable solution probabilities for some instances or lower optimization overhead.
Nevertheless, the proposed method achieves mean solution probabilities above 0.98 for all three instances at the largest values of $p$ considered.
This consistently high final solution probability across the instances studied supports its use as a competitive option when high solution quality is prioritized, and the additional optimization overhead can be accommodated.
}

%%%%%%%%%%%%%%%%%%%%%%%%%%%%%%%%%%%%%%%%%%%%%%%%%%%%%%%%%%%%%%%%%%%%
%%%
%%% Conclusion
%%%
%%%%%%%%%%%%%%%%%%%%%%%%%%%%%%%%%%%%%%%%%%%%%%%%%%%%%%%%%%%%%%%%%%%%
\section{Conclusion}
\label{sec:conclusion}

In this work, we introduced a product-structured variant of the Grover mixer for GM-QAOA, \rev{with independent mixing parameters for each subsystem,} tailored to initial states that admit a product structure over disjoint qubit subsystems.
The proposed approach preserves the search space defined by the initial state as in the original GM-QAOA, while achieving a substantial reduction in circuit complexity by constructing the mixing unitary as a tensor product of local unitaries.
Numerical simulations on the exact-cover problem and the TSP demonstrated that this variant can attain \rev{solution probabilities comparable to those} of the original GM-QAOA, while requiring significantly fewer gates and achieving lower circuit depth.
\rev{These circuit savings are accompanied by increased optimization overhead, reflecting a trade-off between the resources required per circuit and the number of circuit evaluations.
Full-circuit simulations under a depolarizing noise model also support improved robustness to gate noise compared with GM-QAOA.}
We also examined the interplay between constraint encoding strategies and circuit overhead in the context of the TSP, and showed that partial constraint encoding combined with the product-structured mixer \rev{can yield} considerably more compact circuits than full constraint encoding, when compared at the point where comparable solution quality is achieved.
\rev{Comparisons with shared-angle local Grover and partitioned XY-ring mixers further clarified the benefits of independent mixing parameters and the dependence of performance and resource trade-offs on the problem instance.}

Several directions remain open for future work.
First, when gradient-based optimization is employed, the large number of variational parameters increases the number of circuit executions required for gradient estimation.
Thus, developing more efficient parameter optimization strategies or techniques for reducing the number of parameters could further improve performance.
Furthermore, the increased number of variational parameters also raises questions about trainability, and a thorough analysis of barren plateaus in the proposed method remains an important open direction.
In addition, while gradient-based optimization was used in this work to enable a controlled comparison between the original GM-QAOA and ours, gradient-free optimization methods may be more practical in the presence of hardware noise, as they avoid the overhead associated with gradient estimation.
Investigating the performance of the proposed method under such optimizers thus constitutes an important direction for future study.
More broadly, a systematic evaluation of the full optimization process under realistic hardware noise models would provide stronger evidence for the practical advantages suggested by the \rev{resource and noise analyses presented here.}

Furthermore, while the numerical results in this work focus on small instances of the exact-cover problem and the TSP, assessing the scalability of the proposed approach to larger problem instances remains an important open question.
\rev{The linear-ramp evaluations on five- and six-city TSP instances, requiring 16 and 25 qubits, respectively, provide additional evidence that competitive solution probabilities remain accessible within a restricted parameterization beyond the four-city examples.}
Extending the proposed framework to other combinatorial optimization problems with different constraint structures would also be of significant interest.
\rev{The comparison between partial and full constraint encoding suggests that enforcing all constraints through state preparation need not yield the most resource-efficient overall circuit.
Investigating this trade-off beyond the TSP, together with suitable choices of local mixers, is therefore a promising direction.
Further resource and performance comparisons could also examine implementations that combine local Grover, XY, or LX mixers on different subsystems.}
Finally, a rigorous theoretical understanding of the proposed method remains an important open problem.
In particular, explaining why a product-structured mixer can achieve \rev{solution probabilities comparable to those} of the Grover mixer, despite its fundamentally different mixing structure, would provide deeper insight into the design of constraint-aware QAOA algorithms.

\section*{Data Availability}
\rev{
The code used to reproduce the numerical results, together with the data and plotting scripts used to generate the figures, is available at https://github.com/mathcmj/LGM-QAOA.
}

%%%%%%%%%%%%%%%%%%%%%%%%%%%%%%%%%%%%%%%%%%%%%%%%%%%%%%%%%%%%%%%%%%%%%%%%%%%%%%%%%%%%%%%
%%
%% Acknowledgment
%%
%%%%%%%%%%%%%%%%%%%%%%%%%%%%%%%%%%%%%%%%%%%%%%%%%%%%%%%%%%%%%%%%%%%%%%%%%%%%%%%%%%%%%%%
\section*{ACKNOWLEDGMENTS}

This research was supported by the Korea Institute of Science and Technology Information (KISTI) R\&D (Grant No. K26L3M3C3), the National Research Council of Science \& Technology (NST) grant by the Korea government (MSIT) (No. GTL25011-000), and the National Research Foundation of Korea (Grant No. RS-2022-NR068791). This research also was supported using HPC resources and technical support provided by KISTI.

\bibliography{reference}

\end{document}